\documentclass[aps,showpacs,prd,twocolumn,amsmath,amssymb]{revtex4-1}
\usepackage{color}
\usepackage{graphics}
\usepackage{epsfig}
\usepackage{amsmath}
\usepackage{amssymb}
\usepackage{amsthm}
\usepackage[mathscr]{eucal}

\newcommand{\bea}{\begin{eqnarray}}
\newcommand{\eea}{\end{eqnarray}}
\newcommand{\beas}{\begin{eqnarray*}}
\newcommand{\eeas}{\end{eqnarray*}}

\begin{document}

\title{Reply to comment on "Equation of state of dense and magnetized fermion system"}
\author{Efrain J. Ferrer and Vivian de la Incera}
\affiliation{Department of Physics, University of Texas at El Paso, 500 W. University Ave., El Paso, TX 79968, USA}

\maketitle
A Comment \cite{comment} to our recent article \cite{PRC82-065802}, expressed criticism on our estimate of the maximum magnetic field that can exist inside a neutron star and on the pressure anisotropy we found for a magnetized gas of fermions. These two points are relevant for the physics of neutron stars. With our reply we attempt not only to address the Comment's criticisms, but for the sake of the reader's understanding, we also try to clarify the connection between our and some literature's results mentioned in the comment.

\section{Maximum field strength}
As known, the inner magnetic fields of neutron stars are not directly accessible to observation, so their possible values can only be estimated with the help of heuristic methods. A widely accepted estimate of the maximum star's magnetic field $\sim 10^{18} G$ was done in Ref. \cite{Lai-Shapiro} using a macroscopic analysis based on the equipartition of the magnetic and gravitational energies (Virial Theorem), and assuming a constant density and a uniform magnetic field throughout the entire star. As stated in the Comment \cite{comment}, this estimate is in agreement with numerical simulations that used various EoS of nuclear matter \cite{Bocquet, Kiuchi}, although it should be mentioned that the results of Ref. \cite{Bocquet} were later criticized \cite{AstrPhy J 537-2000} because in their calculations the authors omitted the compositional changes in the EoS due to Landau quantization. Same criticism is applicable to Ref. \cite{Kiuchi}, since the magnetic field was not considered in the EoS they used. Ignoring the field in the EoS of nuclear matter is not consistent at strong fields because the EoS is substantially softened by the magnetic field for fields well below the proton critical field $\sim 10^{20}$ G \cite{AstrPhy J 537-2000}.

In Ref. \cite{PRC82-065802} we argued that if the star is a hybrid star, meaning its core is made of deconfined quark matter, while its outer layers are made of nuclear matter, the previous inner field estimates based on a constant mass distribution may not be reliable because in the hybrid star a much larger difference should exist between the matter densities in the outer layers and in the core. In Section IIB of our paper, we did two independent estimates of the maximum magnetic field of a hybrid star assuming a very dense core. The Comment criticism pertains to the first of these estimates. Like in \cite{Lai-Shapiro}, our first estimate was based on the equipartition of the gravitational and magnetic energies in a star of given mass and radius. The main difference with respect to \cite{Lai-Shapiro} was that we considered an ad-hoc model with nonuniform matter and field magnitude distributions. The only purpose we had with this simple example was to illustrate that the maximum field estimate can change just by relaxing the assumption of uniform mass and field distributions that was used in \cite{Lai-Shapiro}. In the second paragraph following Eq.(10) we wrote "we  make no claim that this ad hoc model correctly describes the real way the field varies with the radius in a hybrid star". More importantly, this ad hoc model was not used for any of the main derivations of the paper, so it has no relevance for any of the paper's main results. Notice that there are other attempts in the literature \cite{Pal} of parameterizing the variation of the magnetic field with the stellar radius.

The Comment failed to mention that in \cite{PRC82-065802} we did a second, completely independent, microscopic estimate of the maximum inner field in the last part of Section IIB that not only is model-independent, but it is based on the very natural assumption that the magnetic energy density should be at most as large as the
baryon energy density. What this second estimate indicates is that if the baryon density in the core is large enough to deconfine the quarks (chemical potential $\mu \gtrsim 400 MeV$), the maximum magnetic field there can be as large as $10^{20} G$. Interestingly, this second estimate is similar to the one considered in Ref. \cite{Bocquet}, which is one of the papers cited in the Comment. In \cite{Bocquet}, the criterium to estimate the maximum field at the center was based on the stability condition given by the cancelation between the nuclear fluid pressure and the magnetic pressure, $\Omega_{matter} = B^2/2$,(see Sec. 4.2 in Ref. \cite{Bocquet}), equivalent to the condition $p_\|=0$ of our paper. Now, taking into account that the main contribution to the fluid pressure is $\sim \mu^4$, and that the chemical potential in the case of quark matter can be $\mu \sim 400$ MeV, one straightforwardly obtains $B \sim 10^{20}$ G for the case of quark matter. The smaller field estimate $\sim 10^{18} G$ of \cite{Bocquet, Kiuchi} is connected to the fact that they considered nuclear matter instead of quark matter. An additional confirmation that the condition $\Omega_{matter} = B^2/2$ is reached at field values of higher order than $10^{18} G$ for quark matter, was obtained by numerical calculations in Ref. \cite{Paulucci}, where the EoS of the color superconducting phase of quark matter was investigated.

In summary, the key issue to estimate the maximum inner field is how large the baryon density at the core of a neutron star is, because its magnitude will determine the value of the matter energy density, which in turn sets the upper limit for the magnetic energy density achievable at the core.

\section{Pressure Anisotropy}
In the second section of the Comment \cite{comment}, the authors criticize the anisotropy that arises in the calculation of the parallel and transverse pressures in the presence of a strong magnetic field. According to them, the finding of an anisotropy in the pressure and hence in the EoS of the magnetized system \cite{PRC82-065802, Canuto} is inconsistent, because it ignores the effect of the Lorentz force density related to the magnetization currents. To sustain their claim, the Comment authors cite Ref. \cite{Blandford} and also give two ad hoc examples.

The arguments used in the Appendix of \cite{Blandford} and then repeated, without any new theoretical development, in Section II of \cite{comment}, sustained that the pressure splitting should be eliminated by the creation of an extra term related to the work done by compressing the fermion gas in the transverse direction against the Lorentz force. The new term involves the magnetization current density $(\nabla \times M) \times B$, which would make a contribution, $MB$, to the transverse pressure, hence regaining the pressure isotropy. Neither the Comment, nor Ref. \cite{Blandford} offered a rigorous proof of such a claim, as the exact Lorentz force required for the cancelation was not derived in a consistent way, but introduced by hand.

We highlight that a magnetization current density $\sim \nabla \times M$ is absent in our paper \cite{PRC82-065802}, as well as in \cite{Canuto}, because $\nabla \times M=0$ for the static system with uniform magnetic field we studied. Assuming a uniform magnetic field in our microscopic formulation is well justified, because despite the magnetic field strength of the neutron star may well vary from the core to the surface, the scale length of the field variation in the star is much larger than the microscopic magnetic length \cite{AstrPhy J 537-2000}.

The authors of \cite{comment} also tried to sustain their claim by considering two ad hoc examples that have nothing to do with the problem studied in our paper. The first example consists of a system with boundaries. It not only does not apply to our study of the EoS of the bulk material, but the analysis is done assuming exactly the hypothesis it intends to prove, i.e. that the two "total" pressures should be equal to guarantee the isotropy in the pressures. In the second example, they propose to introduce in the calculation of the thermodynamic potential, the gravitational acceleration, arbitrarily chosen to depend on z, in order to produce an inhomogeneous matter density distribution and hence a nonzero $\nabla \times M$, because $M=-\frac{\partial \Omega}{\partial B}$ would depend on z. This is a completely artificial example, but even worse, it introduces the effects of gravity in $\Omega$, when it is well known that they are insignificant at the microscopic level and hence should be ignored in the EoS. Therefore, none of these examples provides any reliable proof of the Comment's claims.

We point out that the finding of the EoS is based on microscopic calculations that pertain to the properties of the bulk. Energy and pressure can be found in this context as the statistical average of the components of the energy-momentum tensor. The essence of the origin of the pressure anisotropy is contained in the Eq. (81) of our paper \cite{PRC82-065802}. In this equation we wrote the average of the energy-momentum tensor in terms of all the available covariant structures of the system. For example, the breaking of the Lorentz symmetry by the density gives rise to the well-known contribution to the energy, $\mu N$, coming from the second term of (81). In a similar fashion, the presence of the magnetic field, which breaks the spatial rotational symmetry, yields a new tensorial structure reflected in the last term of (81). This last term is responsible for the pressure anisotropy.

To conclude, we underline that Ref. \cite{comment} has not provided any valid proof that the finding of the pressure's anisotropy of our paper is inconsistent. Our result was rigourously proved in \cite{PRC82-065802}. Whether the anisotropy can or not be compensated by the currents generated by the rotation of the star, a surface magnetization produced by boundary effects, etc, is still unclear, but unrelated to the microscopic calculations performed in \cite{PRC82-065802}. To answer this question, a rigorous macroscopic analysis needs to be done, one that uses the equations of magnetohydrodynamics and that is based on a self-consistent treatment of the equations of the stellar structure in a strong magnetic field, i.e. using axisymmetric TOV equations, and the anisotropic EoS of the magnetized matter. No doubt this can be a non trivial, but very interesting and important task. Even though we do not agree with the criticism of our paper raised by Comment \cite{comment} for the reasons explained above, we believe that this discussion may help to call attention to an important question about strongly magnetized stars that still remains unsolved and needs to be investigated self-consistently and in full detail.

{\bf Acknowledgments:} This work has been supported in part by DOE Nuclear Theory grant DE-SC0002179.


\begin{thebibliography}{}
\bibitem{comment} A. Y. Potekhin and D. G. Yakovlev, arXiv: 1109.3783[astro-ph.SR].

\bibitem{PRC82-065802} E. J. Ferrer, et al., Phys. Rev. C \textbf{82}, 065802 (2010).

\bibitem{Lai-Shapiro} D. Lai and S. L. Shapiro, Astrophys. J. \textbf{383}, 745 (1991).

\bibitem{Bocquet} M. Bocquet, et al., Astron.
Astrophys. \textbf{301}, 757 (1995).

\bibitem{Kiuchi}K. Kiuchi and K. Kotake, Mon. Not. R. Astron. Soc. \textbf{385}, 1327
(2008).

\bibitem{AstrPhy J 537-2000} A. Broderick, M. Prakash, and J.M. Lattimer, Astrophys. J. \textbf{537}, 351 (2000).

\bibitem{Pal} D. Bandyopadhyay, S. Chacrabarty and S. Pal, Phys. Rev. Lett. \textbf{79}, 2176 (1997).

\bibitem{Paulucci}L. Paulucci, et al., Phys. Rev.D \textbf{83}, 043009 (2011).

\bibitem{Canuto} V. Canuto and H.-Y. Chiu, Phys. Rev. \textbf{173}, 1210 (1968); \textbf{173},
1220 (1968).

\bibitem{Blandford} R. D. Blandford and L. Hernquist, J. Phys. C \textbf{15}, 6233 (1982).

\end{thebibliography}
\end{document}